\documentclass[useAMS,usenatbib]{mn2e}

\usepackage{graphicx}
\usepackage{url}
\usepackage{bm}
\usepackage{float}
\usepackage{placeins}
\usepackage{setspace}
\usepackage[fleqn]{amsmath}

\title{Disentangling a group of lensed submm galaxies at \textit{z}$\,{\sim}\,$2.9}

\author[T. Mackenzie et al.]{Todd P. MacKenzie$^{1}$,
Douglas Scott$^{1}$, 
Ian Smail$^{2}$,
Edward L. Chapin$^{3}$,\newauthor
Scott C. Chapman$^{4}$,
A. Conley$^{5}$,
Asantha Cooray$^{6,7}$,
James S. Dunlop$^{8}$, \newauthor
D. Farrah$^{9}$,
M. Fich$^{10}$,
Andy G. Gibb$^{1}$,
R. J. Ivison$^{8}$,
Tim Jenness$^{11,12}$,\newauthor
Jean-Paul Kneib$^{13}$,
Gaelen Marsden$^{1}$,
Johan Richard$^{14}$,
E. I. Robson$^{15}$,\newauthor
Ivan Valtchanov$^{16}$,
Julie L. Wardlow$^{17}$
\\
$^{1}$Department of Physics \& Astronomy, University of British Columbia, 6224 Agricultural Road, Vancouver, BC V6T 1Z1, Canada \\
$^{2}$Institute for Computational Cosmology, Department of Physics, Durham University, South Road, Durham DH1 3LE \\
$^{3}$Joint Astronomy Centre, 660 N. A'oh\={o}k\={u} Place, University Park, Hilo, Hawaii 96720, USA\\
$^{4}$Department of Physics and Atmospheric Science, Dalhousie University, 6310 Coburg Rd., Halifax, NS B3H 4R2, Canada \\
$^{5}$Center for Astrophysics and Space Astronomy 389-UCB, University of Colorado, Boulder, CO 80309, USA\\
$^{6}$Department of Physics \& Astronomy, University of California, Irvine, CA 92697, USA\\
$^{7}$California Institute of Technology, 1200 E. California Blvd., Pasadena, CA 91125, USA\\
$^{8}$Institute for Astronomy, University of Edinburgh, Royal Observatory, Blackford Hill, Edinburgh EH9 3HJ, UK\\
$^{9}$Department of Physics, Virginia Tech, VA 24061, USA\\
$^{10}$Department of Physics and Astronomy, University of Waterloo, Waterloo, Ontario, N2L 3G1, Canada\\
$^{11}$Department of Astronomy, Cornell University, Ithaca, NY 14853, USA \\
$^{12}$Joint Astronomy Centre, 660 N. A'ohoku Place, University Park, Hilo, Hawaii, 96720, USA\\
$^{13}$Laboratoire d’astrophysique, Ecole Polytechnique F\'ed\'erale de Lausanne, Observatoire de Sauverny, 1290 Versoix, Switzerland\\
$^{14}$Observatoire de Lyon, F-69561 Saint-Genis-Laval, France \\
$^{15}$UK Astronomy Technology Centre, Royal Observatory, Edinburgh, Blackford Hill, Edinburgh, EH9 3HJ, UK\\
$^{16}$Herschel Science Centre, European Space Astronomy Centre, ESA, 28691 Villanueva de la Ca\~nada, Spain\\
$^{17}$Dark Cosmology Centre, Niels Bohr Institute, University of Copenhagen\\
}

\begin{document}
\maketitle

\begin{abstract}
MS\,0451.6$-$0305 is a rich galaxy cluster whose strong lensing is particularly prominent at submm wavelengths.  We combine new SCUBA-2 data with imaging from {\it Herschel} SPIRE and PACS and {\it HST} in order to try to understand the nature of the sources being lensed.  In the region of the ``giant submm arc,'' we uncover seven multiply imaged galaxies (up from the previously known three), of which six are found to be at a redshift of $z\sim2.9$, and possibly constitute an interacting system.  Using a novel forward-modelling approach, we are able to simultaneously deblend and fit SEDs to the individual galaxies that contribute to the giant submm arc, constraining their dust temperatures, far infrared luminosities and star formation rates.  The submm arc first identified by SCUBA can now be seen to be composed of at least five distinct sources, four of these within the galaxy group at $z\sim2.9$.  The total unlensed luminosity for this galaxy group is $(3.1\pm0.3) \times 10^{12}\,\mathrm{L}_\odot$, which gives an unlensed star formation rate of $(450\pm50)$ M$_\odot$\,yr$^{-1}$.  From the properties of this system, we see no evidence of evolution towards lower temperatures in the dust temperature versus far-infrared luminosity relation for high redshift galaxies.
\end{abstract}


\section{Introduction}

\begin{figure*}
\begin{center}
\includegraphics[width=17cm]{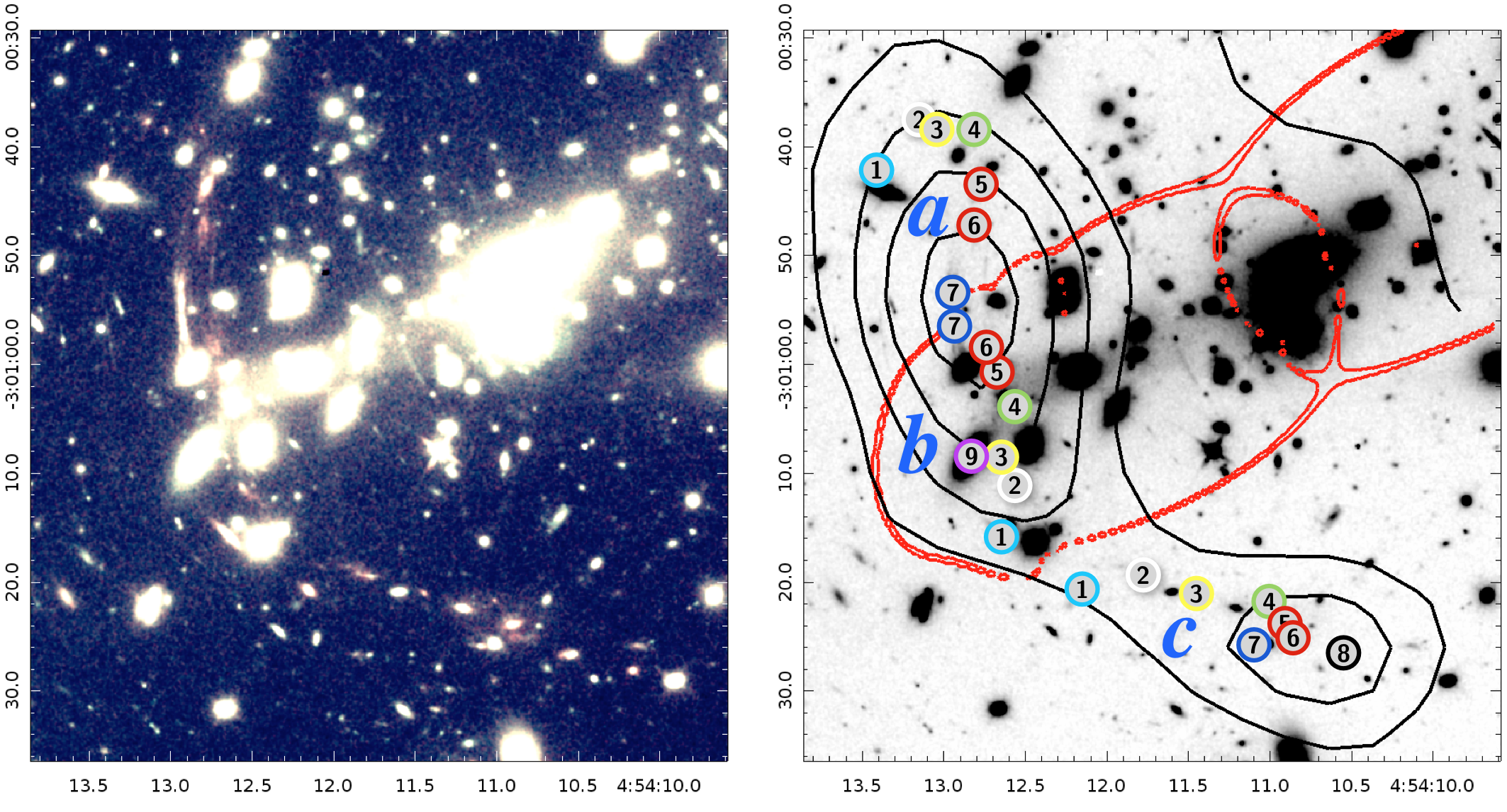}
\newline
\vspace*{0 cm}
\newline
\includegraphics[width=17.5cm]{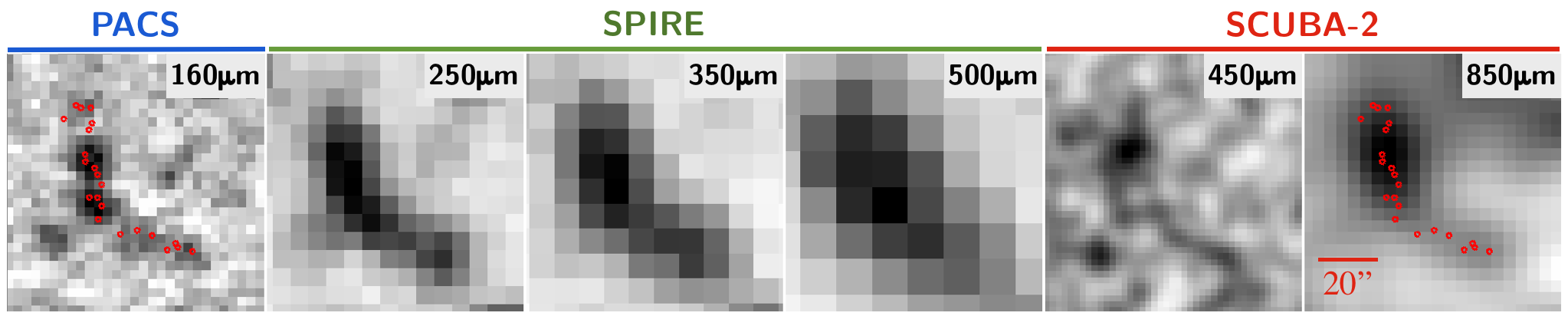}

\end{center}
\caption{{\it Top left}: {\it HST\/} WFC3 colour-composite (red: 1.6\,$\mu$m, green: 1.6 + 1.1\,$\mu$m, blue: 1.1\,$\mu$m), clearly showing the main optical arc (roughly vertical, at about RA$\,{=}\,4^\mathrm{h}54^\mathrm{m}12.9^\mathrm{s}$), offset slightly from the abundance of red images along the submm arc.  The contrast has been stretched to highlight the faint arcs and multiply imaged galaxies.  {\it Top right}: {\it HST} image (1.6 + 1.1\,$\mu$m) with the positions of the multiply-imaged galaxies labelled numerically from 1 through 7, with sub-groups of images labelled as {\it a, b} and {\it c}.  The two EROs and the LBG discovered by \protect\cite{Borys2004} are labelled 5, 6, and 7, respectively.  The red contour denotes the critical line of the lensing model for a redshift of $z=2.911$, while the black contours represent the SCUBA-2 850\,$\mu$m emission.  Galaxy 8 is a singly imaged source with colours similar to those of the other multiply-imaged galaxies and has been found to be important when trying to reproduce the morphology of the submm arc.  Galaxy 9 is a foreground galaxy at a redshift of $z=0.157$.
{\it Bottom}:  The ``giant submm arc'' as seen by {\it Herschel} PACS and SPIRE and SCUBA-2 over more than a factor of five in wavelength range.  The red circles plotted on the shortest and longest wavelength images mark the positions of the galaxies depicted in the top right panel.  It is obvious that this string of multiply imaged $z\sim2.9$ galaxy group sources are responsible for generating the majority of the submm arc.  However, they are too spatially confused for traditional deblending techniques.} 
\label{fig:comparisons}
\end{figure*}

Gravitational lensing has been a useful tool for enabling submm studies.  The first results from SCUBA \cite[e.g.][]{Smail1997}
used ``nature's telescope'' to increase the detection rate of high-redshift submm sources and effectively beat the confusion limit for single
dish studies.  Now
{\it Herschel\/} \citep{herschelobs} has found that lensing is significant for some of the brightest submm sources, with
surveys such as H-ATLAS and HerMES turning up a population of sources which are boosted enough that they can be studied in great
detail in follow-up observations \cite[e.g.][]{Negrello2010,Wardlow2013}.
However, the limited resolution of {\it Herschel}, and of non-interferometric
ground-based observatories such as the James Clerk Maxwell Telescope (JCMT), means that the effects of source blending are a cause of uncertainty in interpreting the results \cite[e.g.][]{karim}, made more difficult in practice, since submm-bright sources are known to be typically merging or interacting systems, where disentangling the contribution to the combined spectral energy distribution (SED) is more complicated still.  Even worse -- while lensing is nominally achromatic, strong lensing of inhomogeneous extended sources within finite beams is {\it not\/} achromatic, since unresolved regions with different spectral properties can be lensed by different amounts.  Thus the existence of strong lensing can be a double-edged sword, boosting the brightness of some sources, but making the detailed interpretation of their spectral energy distributions (SEDs) problematic.  Multi-wavelength studies are key to understanding these complex systems.

MS\,0451.6$-$0305, a massive galaxy cluster at a redshift of 0.55, is lensing several background sources and has been imaged at many different wavelengths: X-ray \citep{Donahue2003}; optical \citep{Gioia1994,Moran2007,Kodama2005,Takata2003}; near-IR \citep{Borys2004,Wardlow2010}; mid-IR \citep{Geach2006}; far-infrared (far-IR) \citep{hermespaper}; mm/submm \citep{first,Borys2004,Wardlow2010}; and radio \citep{Reese2000,Berciano2010}.  In the optical, the previously discovered multiply-imaged sources include an extended optical arc composed of a Lyman-break galaxy (LBG) with a spectroscopic redshift of $z=2.911$, as well as two extremely red objects (EROs) with a redshift of $z=2.9\pm0.1$, determined from lensing models \citep{Borys2004,Berciano2010}.  The two EROs and the LBG are so close in separation ($\sim$10\,kpc in projection) that they potentially constitute an interacting system.  The steep number counts in the submm make lensing much more striking in this waveband than the optical -- at 850\,$\mu$m SCUBA showed a ``giant submm arc,'' by far the brightest feature in this region of the sky, with an extent of around 1 arcminute, consistent with the blending of multiple galaxy images which lie near the critical line in the lensing model.  If the optical galaxies are indeed interacting, the submm arc could be attributed to triggered star formation within one or more of these galaxies.  This scenario is also supported by the radio data, as discussed in \cite{Berciano2010}.  

New observations presented here using the Wide Field Camera 3 (WFC3) on {\it HST}, SCUBA-2 on the JCMT, and PACS and SPIRE on {\it Herschel}\footnote{{\it Herschel} is an ESA space observatory with science instruments provided by European-led Principal Investigator consortia and with important participation from NASA.}, shed new light on what is generating the submm arc.  With the deeper {\it HST} images and a new {\sc Lenstool} \citep{lenstool1,lenstool2,lenstool3} lensing model, we have discovered {\it seven} multiply-imaged galaxies (including the LBG and two EROs) in the region of the submm arc.  Six of these multiply-imaged galaxies are consistent with a redshift of $z\sim2.9$ and probably constitute an interacting galaxy group.  To properly analyse the submm imaging of SCUBA-2 and {\it Herschel}, we have developed a new approach to disentangle the confused components generating the submm arc, which fully exploits the multiply-imaged and differentially-magnified nature of the system, and allows us to directly estimate both the dust temperature, $T_\mathrm{d}$, and the far-infrared luminosity, $L_{\mathrm{IR}}$, (and thus star formation rate, SFR) for each of the contributing galaxies.  This allows us to probe the $T_\mathrm{d}$ verus $L_{\mathrm{IR}}$ relation for intrinsically less luminous galaxies at high-$z$ than traditional blank field surveys.  Possible evolution of this relation with redshift allows us to probe the properties of star formation in the early Universe \citep[e.g.][ and Smail et al. in press]{chapman2002,chapman2005,pope2006,kovacs2006,chapin2011,symeonidis,swinbank,sklias}.  Our method significantly improves upon the conventional method of extracting sources, or smoothing and binning multi-wavelength data to the worst resolution, before fitting SEDs (a process that destroys useful information).

This paper is organised as follows.  In \S~\ref{lenstoolmodel} we introduce the {\it HST} optical data and the lensing model.  In \S~\ref{scuba2data} we present the SCUBA-2 data and in \S~\ref{herschel} the {\it Herschel} data.  In \S~\ref{sedmodelling} we present the SED model and image reconstruction methods and in \S~\ref{modelfitting} the model fitting procedure.  \S~\ref{section:results} discusses the results and \S~\ref{conclusions} finishes with the conclusions.  Throughout we employ a $\Lambda$CDM cosmology with $\Omega_\Lambda=0.7$, $\Omega_\mathrm{m}=0.3$, and $H_0=70\,\mathrm{km}\,\mathrm{s}^{-1}\,\mathrm{Mpc}^{-1}$.

\section{\textbf{\emph{HST}} and the Lensing Model}
\label{lenstoolmodel}
Although the main motivation for our study comes from the new submm data, it makes the most scientific sense to first describe the optical data.
We retrieved previously unpublished observations using WFC3 on {\it HST} from the Canadian Astronomical Data Centre (program 11591).  The observations were taken at 1.1 and 1.6\,$\mu$m with 2400 and 2600 second exposures, respectively.  A  small pointing shift in the data, with respect to {\it HST} data published by \cite{Borys2004} and \cite{Berciano2010}, was corrected by aligning to the older {\it HST} data in this field.  These observations reveal a host of new red objects in the region of the submm arc (Fig.~\ref{fig:comparisons}).

\begin{figure*}
\begin{center}
\includegraphics[width=17.5cm]{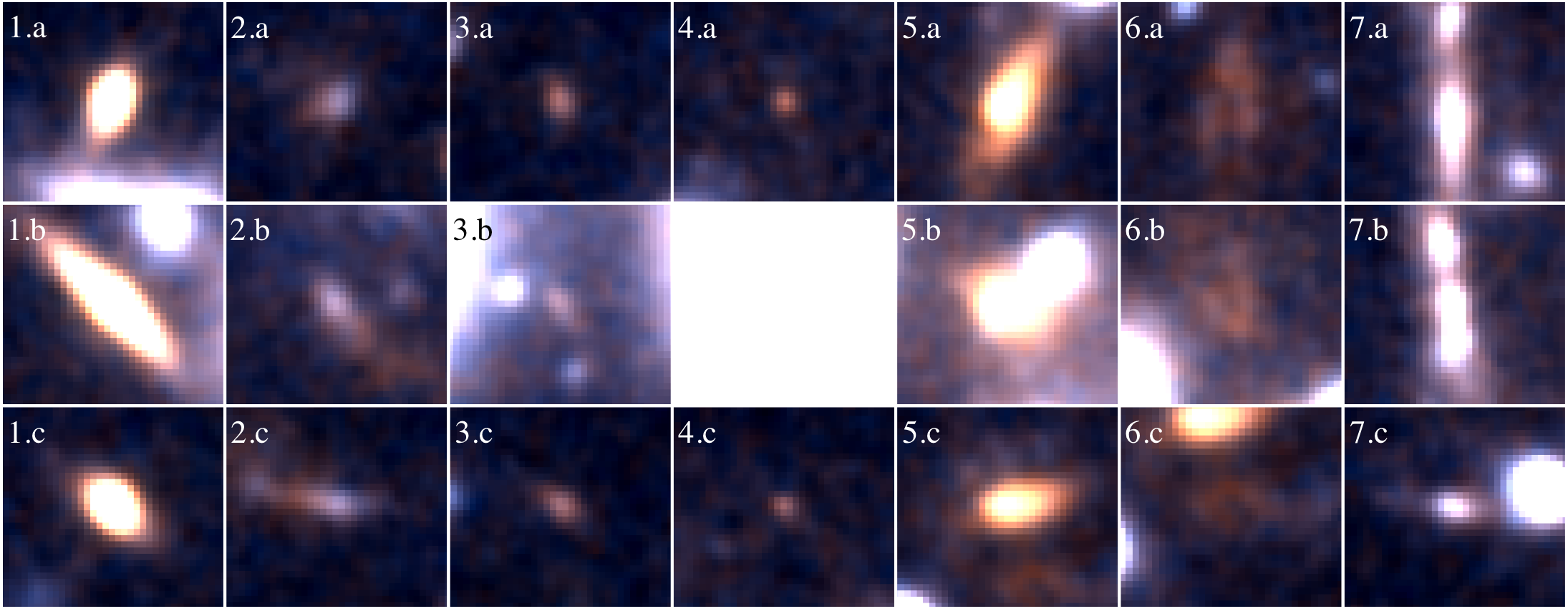}
\end{center}
\caption{{\it HST} cut-outs at the locations of the seven multiply imaged galaxies listed in Table~\ref{images} with each column displaying the multiple images of a single galaxy.  The letters refer to the three sub-groups of images labelled in Fig.~\ref{fig:comparisons}.  Image 4.b is not shown because it is obscured by foreground galaxies.}
\label{fig:zoomed}
\end{figure*}

Using {\sc Lenstool} \citep{lenstool1,lenstool2,lenstool3} and a new lensing model for the cluster, we were able to identify {\it four} new multiply-imaged galaxies within the {\it HST} images in the region of the submm arc.  Table~\ref{images} lists the positions, amplifications factors, and redshift estimates derived from our model for of each of the seven multiply-imaged galaxies within the region of the submm arc.  
Fig.~\ref{fig:comparisons} shows the dramatic positional arrangement of the multiple images with respect to the ``giant submm arc'' and the available submm data.  Enlarged cut-outs of the multiply imaged galaxies are shown in Fig.~\ref{fig:zoomed}.
\cite{Borys2004} have already suggested that Galaxies 5, 6 and 7 are likely to be an interacting group at $z\sim2.9$.  Our new model confirms their analysis and adds Galaxies 2, 3, and 4 to the same group, expanding it to a group of six galaxies at $z\sim2.9$.  Galaxy 1 is found to have a slightly higher redshift of $z=3.11\pm0.03$ derived from the lensing model, and thus not likely associated with the interacting group.  

Galaxy 8 is not multiply imaged, but has similar colours to the rest of the multiply-imaged galaxies and has a disturbed morphology.  If it is at the same redshift as the interacting group, our lensing model predicts no multiple images, and thus we have no constraints on its redshift from the lensing model.  However, we have found that submm emission originating from near its position is important for reproducing the morphology of the submm arc (see \S~\ref{section:results}), and thus we have included it in our model (see \S~\ref{model}). 

Galaxy 9 is a foreground galaxy at $z=0.157$ and has associated MIPS 24\,$\mu$m (not described here) and PACS emission (see \S~\ref{herschel}), thus is also included in our model as a possible source of submm emission.

It is apparent that the nature of the submm arc is significantly more complicated than previously thought and is likely a combination of several of the galaxies described above.  More details concerning the {\sc Lenstool} modelling will be presented in a forthcoming paper by Kneib \& Richard (in prep.).

\section{New Submm Imaging}
\subsection{SCUBA-2}
\label{scuba2data}

The cluster was observed with SCUBA-2 \citep{scuba2} on the JCMT during commissioning, as part of ``Guaranteed Time'' for the instrument team.  Since the submm arc had already been observed at 850\,$\mu$m using SCUBA \citep{Borys2004}, the motivations for the new observations were: (1) to confirm the bright lensed structure with SCUBA-2, without the complications introduced by SCUBA's requirement to chop \citep{Borys2004}; and (2) to detect the lensed structure at 450\,$\mu$m, at a resolution better by about a factor of two, with the hope of resolving the submm arc into individual sources.  The data were reduced using a configuration file optimized for blank fields using the {\sc smurf} data reduction software for SCUBA-2 \citep{smurf}.

At 850\,$\mu$m, the submm arc is detected at high signal-to-noise by SCUBA-2 (see Fig.~\ref{fig:comparisons}).  Its brightest part is elongated roughly north-south, and at the southern end curves to the west, just as in the original SCUBA image. The higher-resolution 450\,$\mu$m data trace a largely similar structure, but at a lower relative sensitivity, with a signal-to-noise ratio of about 3 after smoothing with the beam, for the brightest portion of the lensed emission.  
The SCUBA-2 data are constrained by both resolution at 850\,$\mu$m and sensitivity at 450\,$\mu$m, and thus only limited conclusions can be obtained from these two channels alone.  Fig.~\ref{fig:comparisons} shows the SCUBA-2 data alongside the {\it Herschel} SPIRE and PACS images for comparison, while Fig.~\ref{fig:comparisons} shows smoothed 850\,$\mu$m contours plotted over the {\it HST} imaging.

\subsection{\textbf{\emph{Herschel}}}
\label{herschel}

Confusion-limited images of MS\,0451.6$-$0305 using {\it Herschel} SPIRE \citep{spire, swinyard} were taken as part of the guaranteed time program HerMES \citep[the {\it Herschel} Multi-tiered Extragalactic Survey,][]{hermespaper}.  The cluster was imaged at the three SPIRE wavelengths of 250, 350 and 500\,$\mu$m with FWHM beam sizes of 18.1, 24.9 and 36.2 arcseconds, respectively \citep{spire}.  A detailed description of the map-making procedure is given in \cite{hermesmap1}, and the most recent updated method described in \cite{hermesmap2}. To ensure accurate astrometry, we have stacked on the positions of over 900 {\it Spitzer} MIPS 24\,$\mu$m sources that overlap with the field and have corrected a 1.3 arcsecond shift in RA and 0.4 arcsecond shift in Dec.  The uncertainty in this correction is 0.2 arcseconds, calculated by bootstrapping the 24\,$\mu$m source list.

Two PACS \citep{pacs} observations taken as part of the PACS Evolutionary Probe key program \citep{pacsdata} are also available and were processed using the ``multiple obsid scanMapDeepSurvey'' pipeline within {\sc HIPE} 10 \citep{hipe}. The default units were converted from Jy\,pixel$^{-1}$ to Jy\,beam$^{-1}$ by multiplying by the beam area and dividing by the pixel area.  The beam area for the 160\,$\mu$m point spread function (PSF) was found to be 180 arcsec$^2$ and was computed by integrating over the beam profile provided by the NASA {\it Herschel} Science Center.  The FWHM at 160\,$\mu$m is 11.6 arcseconds.  
For galaxies at $z\sim3$, 70\,$\mu$m PACS data are expected to be dominated by warm dust, which is not well reproduced by the simple SED model adopted in \S~\ref{model}, and are therefore not used in this study.

The submm arc is detected across all the available submm bands (see Fig.~\ref{fig:comparisons}), but with the large number of multiply-imaged galaxies (seen in Fig.~\ref{fig:comparisons}) that are strung along the submm arc, it is unclear which galaxies are contributing.  The morphology of the submm arc seen in each image is a function of both the telescope PSFs and the SEDs of the contributing galaxies.  In addition to determining which galaxies are contributing, we would also like to constrain their physical properties.  With the lensing model well constrained by the {\it HST} observations (see \S~\ref{lenstoolmodel}) and this wealth of multi-wavelength data, it is clear that a comprehensive modelling approach is required.

\section{A framework for fitting SEDs to confused counterparts}
\label{model}

Both \cite{Borys2004} and \cite{Berciano2010} performed limited modelling of the optical and radio counterparts, respectively, in an attempt to reproduce the observed submm arc.  Their approach of smoothing different plausible components with the SCUBA 850\,$\mu$m beam showed that the LBG and two EROs are likely contributors, but neither could fully reproduce the observed submm arc.  With new SCUBA-2 and {\it Herschel} observations, we are able to expand on this approach and have developed a framework for fitting SEDs to the confused optical counterparts, fully exploiting the strong gravitational lensing of this system.

While source plane reconstruction of multiply-imaged galaxies is an effective approach for high-resolution imaging \citep[e.g.][]{reconstruct1,reconstruct2}, it fails in the confused regime.  Because the galaxies blend together in the submm, it is impractical to trace photons back through the lensing potential and into the source plane, since much of the photon positional information has been lost due to the large telescope beams.  Instead, we use the high-resolution {\it HST} imaging to identify candidate counterparts to the submm galaxies in the optical, and use their positions as priors for the origin of any submm emission.  We then forward-model the galaxy SEDs through the telescope filters, and use the amplification factors derived from the lensing model for each galaxy image, to reproduce the submm arc in each wavelength channel separately.  Essentially, we are fitting SEDs of galaxies directly to the data, without the need for first deblending and extracting sources or smoothing and rebinning our data to the worst resolution (a process that destroys useful information).

Our method is complementary to that employed by \cite{fu}, where they forward-model a single submm source through the gravitational lens, allowing the position to vary, to reproduce the observed morphology of their SMA and VLA observations.  With their model, they were able to show that the source of the gas and dust emission was offset from the optical counterpart.  However, the gravitational lensing in this case is galaxy-galaxy lensing and the observations have much higher resolution than either the SCUBA-2 or {\it Herschel} observations presented here.  The gravitational lensing presented here is for a group of galaxies being lensed by a foreground cluster and thus the set of multiple images subtends a much larger area on the sky than galaxy-galaxy lensing.  The optical imaging provides positions which are more than adequate for our purposes, since, with the resolutions of SCUBA-2 and {\it Herschel}, any small offset of the submm emission from their optical counterparts will not have a strong effect on the morphology of the submm arc;  the strongest effect of an offset would be seen in the relative amplifications of the multiple images.  Our method is novel in that we reproduce the morphology of the submm emission across multiple wavelengths, while simultaneously fitting source SEDs, thus tying together the multi-wavelength data.  These two complementary techniques (detailed source plane reconstruction and forward modelling SEDs at fixed source positions) could be combined in the future, given the proper observations.

\subsection{Model SED and image reconstruction}
\label{sedmodelling}

The first ingredient we need is an SED model for our galaxies in the submm.  For the longer wavelength channels of SCUBA-2 and {\it Herschel} SPIRE, the SED of a galaxy is well represented by a modified blackbody with a single temperature:

\begin{multline}
S_{i}(\nu,T_\mathrm{d}, \beta, z, C) = \\ C\left(\frac{\nu (1+z)}{\nu_0}\right)^{\beta} \Big(\nu(1+z)\Big)^3 \left[\mathrm{exp}\left(\frac{h \nu (1+z)}{k_\mathrm{B} T_{\mathrm{d}}}\right)-1\right] ^{-1},
\end{multline}

\noindent
where $S_{i}$ is the flux density of galaxy $i$, $C$ is a normalization factor, $\nu$ is the observed frequency, $\nu_0 = 1.2\,{\rm THz} = c /(250\,\mu{\rm m})$, $\beta$ is the dust emissivity, $T_{\mathrm{d}}$ is the dust temperature, and $z$ is the redshift.  Due to the high redshifts of our galaxies, the shorter wavelength channels of {\it Herschel} are dominated by hot dust and are better represented by a power law on the Wien side, i.e.,

\begin{equation}
S_{i}(\nu,T_{\mathrm{d}}, \beta, z, C)\propto \nu^{-\alpha}, 
\end{equation}

\noindent
where the power law amplitude and the frequency at which to switch between the power law and modified blackbody are chosen so that the transition is smooth (i.e. the two functions and their first derivative are continuous);  such a model has been used by \cite{powerlaw}, for example.  We then propagate the individual galaxy SEDs through each telescope bandpass filter:

\begin{equation}
\bar{S}_{b,i}(T_{\mathrm{d}}, \beta, z, C) = \frac{\int S_{i}(\nu,T_{\mathrm{d}}, \beta, z, C) T_b(\nu)\mathrm{d}\nu}{\int T_b(\nu)f_b(\nu)\mathrm{d}\nu},
\end{equation}

\noindent
where $\bar{S}_{b,i}$ is the flux density averaged over channel $b$ for galaxy $i$, $T_b(\nu)$ is the transmission for channel $b$, and $f_b(\nu)$ is a calibration factor.  For {\it Herschel}, $f_b(\nu)=\nu_0/\nu$, due to assuming a power law SED shape for observed sources, where $\nu_0$ is equal to $c /(160\,\mu{\rm m})$, $c /(250\,\mu{\rm m})$, $c /(350\,\mu{\rm m})$, and $c /(500\,\mu{\rm m})$ for 160, 250, 350 and 500\,$\mu$m, respectively.  We assume a constant calibration factor, $f_b(\nu)=1$, for SCUBA-2, since the bandpass filters are relatively narrow and we are firmly on the Rayleigh-Jeans side of the spectrum.  

Using the lensing model, the SCUBA-2 and {\it Herschel} images are reconstructed as follows:

\begin{equation}
M_{b}(\boldsymbol{x})=\sum\limits_i \sum\limits_j A_{ij} \bar{S}_{b,i}(T_{\mathrm{d}}, \beta, z, C) P_\nu(\boldsymbol{x}-\boldsymbol{r}_{ij}) + B_b.
\end{equation}

\noindent
Here $M_{b}(\boldsymbol{x})$ is the flux at position $\boldsymbol{x}$ for frequency channel $b$, $A_{ij}$ is the amplification factor for image $j$ of Galaxy $i$ derived from the lensing model, $P_\nu(\boldsymbol{x}-\boldsymbol{r}_{ij})$ is the response function (i.e. the telescope beam), with $\boldsymbol{r}_{ij}$ denoting the position of image $j$ of Galaxy $i$, and $B_b$ is the image background.

The response functions for the {\it Herschel} SPIRE channels are approximated as Gaussians with FWHM of 18.1, 24.9 and 36.2 arcseconds at $250$, $350$ and $500$\,$\mu$m, respectively, and 11.6 arcseconds at 160\,$\mu$m for {\it Herschel} PACS.  Due to the high-pass filtering of the SCUBA-2 data, we need to ensure that we have an accurate model of the effective response function, thus we simulate 7 and 15 arcsecond FWHM point-sources, for the 450 and 850\,$\mu$m data, respectively, within the {\sc smurf} data-reduction software, and approximate the effective response function by fitting double Gaussians to their resulting shapes.  The result for the 450\,$\mu$m response function is a Gaussian with FWHM of 6.86 arcseconds and an amplitude of 0.893, plus a second Gaussian with FWHM of 34.6 arcseconds and amplitude of $-$0.015.  The result for the 850\,$\mu$m response function is a Gaussian with FWHM of 13.9 arcseconds and a amplitude of 0.869, plus a second Gaussian with FWHM of 25.9 arcseconds and amplitude of $-$0.077.  

\subsection{Model fitting}
\label{modelfitting}

The model is fit to the data using an MCMC Metropolis-Hastings algorithm \citep{mcmc1,mcmc2} with Gibbs sampling \citep{gibbs}.  Since the SCUBA-2 450 and 850\,$\mu$m and PACS 160\,$\mu$m data are limited by instrumental noise, the $\log$ likelihood functions for these data are calculated as follows:

\begin{equation}
-\log L_{b}=X_{b}+ \sum\limits_k\frac{(D_{b}(\boldsymbol{x}_{k})-M_{b}(\boldsymbol{x}_{k})/c_b)^2}{2\sigma_{b,k}^2},
\end{equation}

\noindent
where subscript $b$ denotes the band, $D_{b}(\boldsymbol{x}_{k})$ are the data, $\boldsymbol{x}_{k}$ denotes the position of pixel $k$ in the image, $\sigma_{b,k}^2$ is the instrumental noise for pixel $k$, $c_b$ is the instrument calibration factor (with a mean value of unity), and $X_{b}$ is a constant.

The $\log$ likelihood function for the {\it Herschel} SPIRE data is more complicated, since we are limited by extragalactic confusion noise as opposed to instrumental noise.  The confusion limit in each channel is 5.8, 6.3 and 6.8\,mJy at 250, 350 and 500\,$\mu$m, respectively \citep{confusion}.  This means that the residuals after subtracting the model will be: (i) much larger than instrumental noise; (ii) correlated spatially with the beam; and (iii) correlated across wavelengths.  This is because confusion noise is real signal generated from many faint sources that are all blending together to produce an unknown and correlated variable background.  Taking confusion into account, the $\log$ likelihood function for the {\it Herschel} SPIRE data is therefore

\begin{equation}
-\log L_{\mathrm{SPIRE}}=X_{\mathrm{SPIRE}}+ \frac{1}{2}\bm{R}^\mathrm{T}\bm{C}^{-1}\bm{R},
\end{equation}

\noindent
where $\bm{R}$ is a one-dimensional list of the residuals, and contains all three channels of SPIRE data ($\bm{R}=\{D_{250}(\boldsymbol{x}_{k})-M_{250}(\boldsymbol{x}_{k})/c_{250},D_{350}(\boldsymbol{x}_{k})-M_{350}(\boldsymbol{x}_{k})/c_{350},D_{500}(\boldsymbol{x}_{k})-M_{500}(\boldsymbol{x}_{k})/c_{500}\}$), and $\bm{C}^{-1}$ is the inverse covariance matrix for the residuals.  The covariance matrix, $\bm{C}$, is estimated using the GOODS-North HerMES field, also observed with {\it Herschel} SPIRE.  This is the largest blank {\it Herschel} field with instrumental noise similar to that of the MS\,0451.6$-$0305 data, and has an area of 0.1 deg$^2$.  To estimate the covariance matrix, we extract cut-outs from the GOODS-North field, with the same dimensions as the MS\,0451.6$-$0305 data, and calculate the covariance between all the pixels.  We then average the covariance matrices of each set of cut-outs to obtain an estimate of the true covariance matrix, ignoring regions with standard deviations greater than twice the confusion limit in any channel to avoid regions with significantly bright sources.  The total $\log$ likelihood is then

\begin{equation}
\log L_{\mathrm{total}}=\log L_{850}+\log L_{450}+\log L_{\mathrm{SPIRE}}+\log L_{160}.
\end{equation}

Flux calibration uncertainties, $c_b$, are taken into account during the fitting procedure by setting priors on $c_b$ for each band.  The flux calibrations of the 160, 450 and 850\,$\mu$m data are 5\%, 2.5\% and 5\%, respectively \citep{pacscalibration, scuba2calibration}.  SPIRE waveband calibrations are correlated, with a covariance matrix

\begin{equation}
C=\left| \begin{array}{ccc}
  0.001825 & 0.0016 & 0.0016 \\
  0.0016 & 0.001825 & 0.0016 \\
  0.0016 & 0.0016 & 0.001825 
  \end{array}\right|, 
\end{equation}

\noindent
where the calibration is normalised to unity \citep[][, 4\% correlated uncertainty between bands plus 1.5\% uncorrelated between bands]{bendo}.  The calibration uncertainties are a small effect when compared to the instrumental and confusion noise within the observations.

We set no prior on the amplitude of the modified blackbodies or each unknown image background. A hard prior, $T>10\,\mathrm{K}$, is motivated by the fact that neither \cite{kingfish} nor \cite{herschelSEDpaper} found any colder {\it Herschel\/} galaxies in either the nearby or distant Universe, respectively.  We have virtually no constraining power on the dust emissivity, $\beta$, and we therefore fix it to a nominal value of 1.5.  We fix $\alpha$ to 2.0 as found by \cite{Casey2012}.
While we have corrected the relative pointing of {\it Herschel\/} and {\it HST}, we are unable to find any significant pointing shift in the JCMT due lower signal-to-noise in the map than what is available in the {\it Herschel} SPIRE observations.  The nominal pointing accuracy is 1.5 arcseconds, and thus we include this as a prior and marginalise over any possible pointing offset along with the image backgrounds.


Table~\ref{images} lists the possible contributing galaxies in our model.  This consists of the seven multiply-imaged galaxies, one singly-imaged red galaxy (Galaxy 8) with disturbed morphology, and one foreground galaxy (Galaxy 9) with associated MIPS 24\,$\mu$m and PACS 160\,$\mu$m emission.  This brings the total to nine possible contributing galaxies. Their positions are derived from the {\it HST} data and their amplification factors are derived from the lensing model.  The redshift of Galaxy 8 is set to a nominal value of $z=2.9$ and we report the lensed far-IR luminosity and SFR for this galaxy.  Galaxy image positions, amplification factors and redshifts are held fixed during the fitting procedure since their uncertainties are small.

Far-IR luminosities are calculated by integrating the rest-frame SEDs from 8 to 1000\,$\mu$m and SFRs are calculated using the relation measured by \cite{sfrconversion}:

\begin{equation}
\mathrm{SFR}=1.49\times10^{-10}\mathrm{M}_\odot \mathrm{yr}^{-1} L_{\mathrm{FIR}}/\mathrm{L}_\odot.
\end{equation}

\noindent
When reporting the uncertainties in our SFR values for each galaxy in our model, we consider only the uncertainty in far-IR luminosity and do not include any uncertainty in this empirical relationship.

\section{Results and Discussion}
\label{section:results}
\begin{figure}
\begin{center}
\includegraphics[width=7cm]{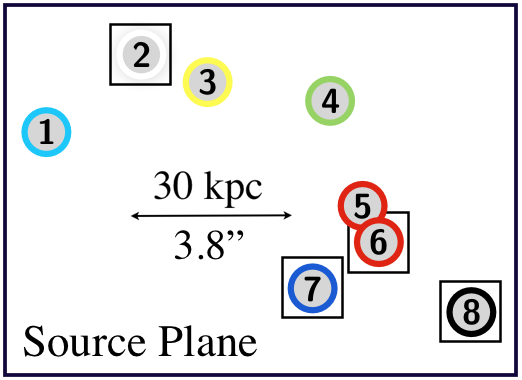}
\end{center}
\caption{Source plane arrangement of the $z\sim2.9$ group galaxies.  Galaxies 2 through 7 are consistent with being at this redshift.  Galaxy 1 lies at a slightly higher redshift, while Galaxy 8 is assumed to be part of the $z\sim2.9$ group.  The boxes highlight the galaxies found to be generating the majority of the submm arc.  These sources are so close to each other that they are likely to be interacting.}
\label{sourceplane}
\end{figure}

\begin{figure*}
\begin{center}
\includegraphics[width=16cm]{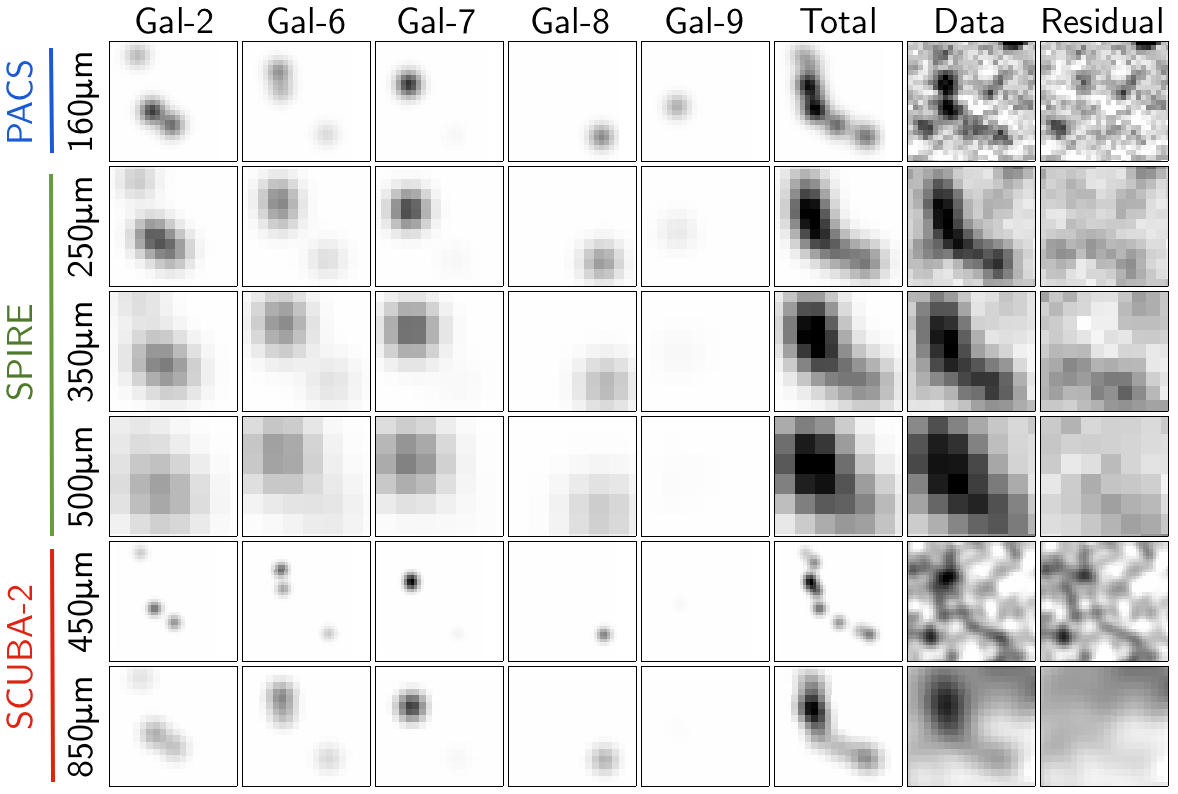}
\end{center}
\caption{Decomposition of the submm arc into each contributing galaxy for the best-fit model, the total emission for the best-fit model, the data, and the residual after subtracting the model from the data.  The columns display the contributions for individual galaxies accross the six wavelength channels.  Due to the differential amplification and unique positions of the multiple images, the emission from each galaxy is morphologically unique and this is what enables us to disentangle their contributions.  The data and residual components for the SCUBA-2 channels have been smoothed with the FWHM for each respective wavelength.  The pixel sizes are 3, 6, 8.3, 12, 2, and 4 arcseconds at 160, 250, 350, 500, 450, and 850\,$\mu$m, respectively.}
\label{fig:components}
\end{figure*}

\begin{figure*}
\begin{center}
\includegraphics[width=17.5cm]{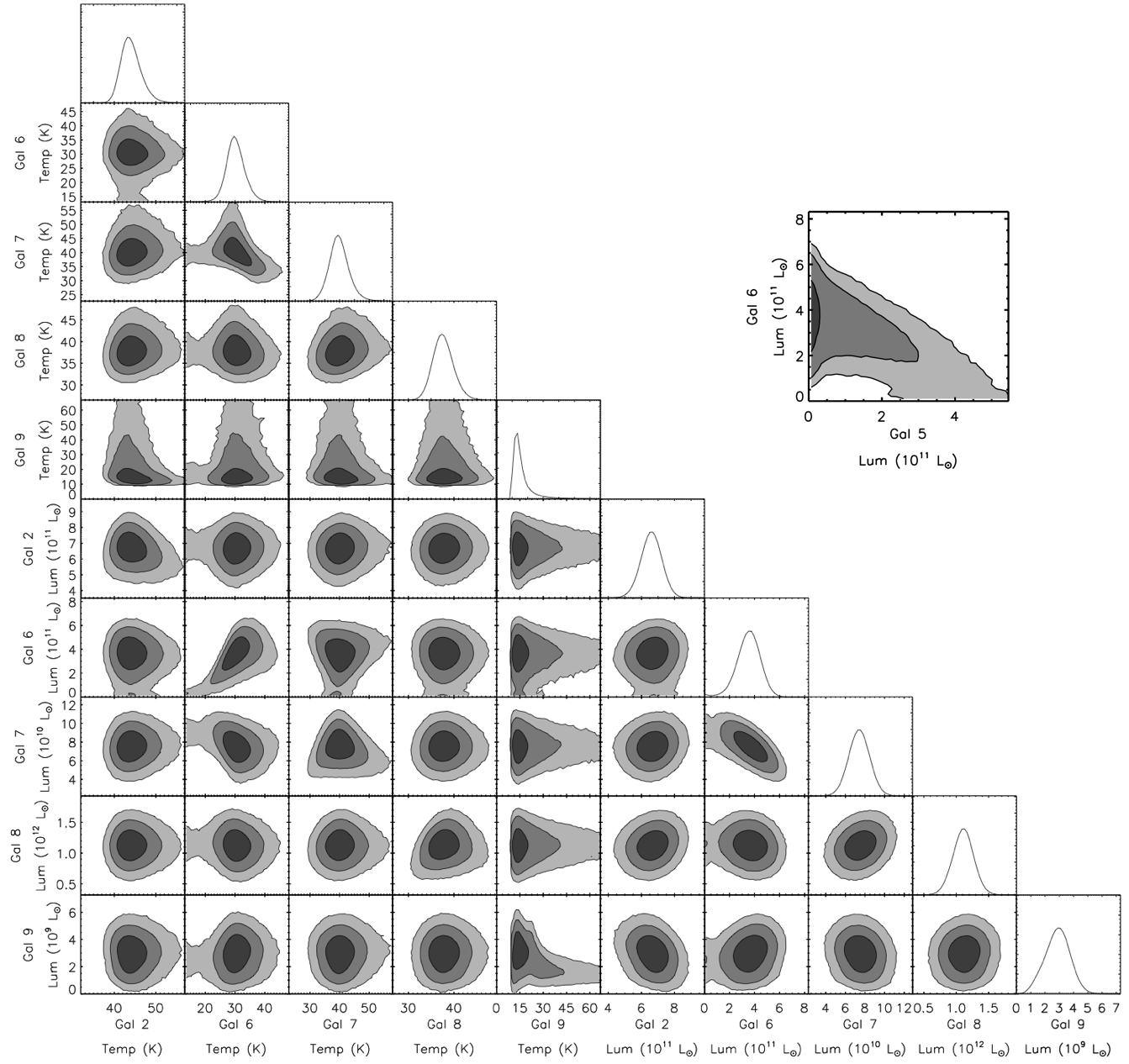}
\end{center}
\caption{MCMC likelihood contours for temperature and far-IR luminosity for the galaxies that were found to contribute to the submm arc.  The contour levels are 68\%, 95\% and 99.7\% confidence intervals.  Because of the morphological uniqueness of the lensing for each individual galaxy, there are few degeneracies here, despite the images of the system being spatially confused.  The most obvious degeneracy is between the far-IR luminosity of Galaxy 6 and Galaxy 7.  {\it Top right}: The likelihood contours for the model show a degeneracy between Galaxy 5 and Galaxy 6 in far-IR luminosity.  Galaxy 5 has associated radio emission, but it exceeds that expected from SFR alone, thus suggesting an AGN component.}
\label{fig:triangleplot}
\end{figure*}

All galaxies listed in Table~\ref{images} are included in our model and when fitted, we can clearly identify Galaxies 2, 6, 7, 8, and 9 as the sources of submm emission generating the submm arc.  Fig.~\ref{sourceplane} shows the positional arrangement of the $z\sim2.9$ galaxy group in the source plane with squares highlighting the galaxies responsible for generating the majority of the submm arc.  Fig.~\ref{fig:components} shows the data, the best-fit model, and the residuals after subtracting the model from the data.  Also included in the figure is a decomposition of the submm arc into the unique contributions of each galaxy to the total best-fit model.  Fig.~\ref{fig:triangleplot} shows the MCMC likelihood contours for temperature and far-IR luminosity for these five galaxies, and Table~\ref{results} lists the results along with SFRs and upper limits for Galaxies 1, 3, 4, and 5.  It is apparent in the MCMC likelihood contours that there is a strong degeneracy between the far-IR luminosities of Galaxies 6 and 7.

Fig.~\ref{fig:triangleplot} shows a degeneracy between the luminosity of Galaxy 6 versus Galaxy 5 in our model, due to their close proximity.  While our model prefers emission from Galaxy 6, \cite{Berciano2010} found that Galaxy 5 has associated radio emission, and hence we might consider that the submm emission attributed to Galaxy 6 in our model actually originates from Galaxy 5.  We can test this hypothesis using the far-IR-to-radio correlation to predict a luminosity for Galaxy 5 and by also removing Galaxy 6 from our model and perform the fitting procedure again (thus forcing our model to attribute a portion of its luminosity to Galaxy 5), and then comparing the results.  When doing so, we find that Galaxy 5 is attributed a luminosity of $(4.5\pm0.9)\times 10^{11}\,\mathrm{L}_\odot$ by our model (i.e essentially all the luminosity of Galaxies 5 and 6 together).  Using the peak flux density measurements of \cite{Berciano2010} at 1.4\,GHz and the amplification factors in Table.~\ref{images}, the unlensed 1.4\,GHz flux density for Galaxy 5 is $(11\pm1)\,\mathrm{\mu Jy}$.  With these two measurements, we can calculate the logarithmic ratio of the far-IR flux to radio flux density, $q_{\mathrm{IR}}=\log_{10}[(S_{\mathrm{IR}}/3.75\times 10^{12}\,\mathrm{Wm}^{-2})/(S_{1.4}/\mathrm{Wm}^{-2}\mathrm{Hz}^{-1})]$.  We assume a power law for the radio SED, $S_{\mathrm{radio}}\propto\nu^\alpha$, with $\alpha=-0.8$, and we K-correct for redshift.  We find $q_{\mathrm{IR}}=1.67\pm0.09$, which is 2-$\sigma$ below the relation found by \cite{herschelradio} for high-$z$ galaxies, $q_{\mathrm{IR}}=2.3\pm0.3$.  This indicates that Galaxy 5 may have excess radio emission, suggesting contribution from an AGN, rather than radio emission associated with star formation.  For this reason, we tend to follow the results which come from our model fitting, i.e. that Galaxy 6 dominates the far-IR emission.  Nevertheless, it remain the case that interpretation of this pair is difficult with existing data.

Using ALMA to obtain high-resolution imaging, \cite{alma} recently showed that many of the submm galaxies (SMGs) previously detected in the LABOCA ECDFS Submillimeter Survey (LESS) are in fact composed of multiple fainter sources.  The group of galaxies behind MS\,0451.6$-$0305, consisting of Galaxies 2 through 8, is another good example of SMGs being composed of several sources.  Unlensed, this $z\sim2.9$ group would appear as a point source to any of the current single-dish submm telescopes, with flux densities of $3.8\pm0.5$~mJy, $8.5\pm0.9$~mJy, $10.4\pm1.1$~mJy, $8.0\pm0.9$~mJy, $8.9\pm1.0$~mJy, and $2.5\pm0.3$~mJy at 160, 250, 350, 500, 450 and 850\,$\mu$m, respectively.  This would put the group below the LESS survey threshold of 4.44\,mJy at 870\,$\mu$m, hence we are seeing evidence of submm source multiplicity extending to the fainter flux densities.  The coincidence of being highly magnified by a massive foreground cluster allows us to study this group in much greater detail than would otherwise be possible, but we cannot tell how rare such SMG groups might be.

\begin{figure}
\begin{center}
\includegraphics[width=8cm]{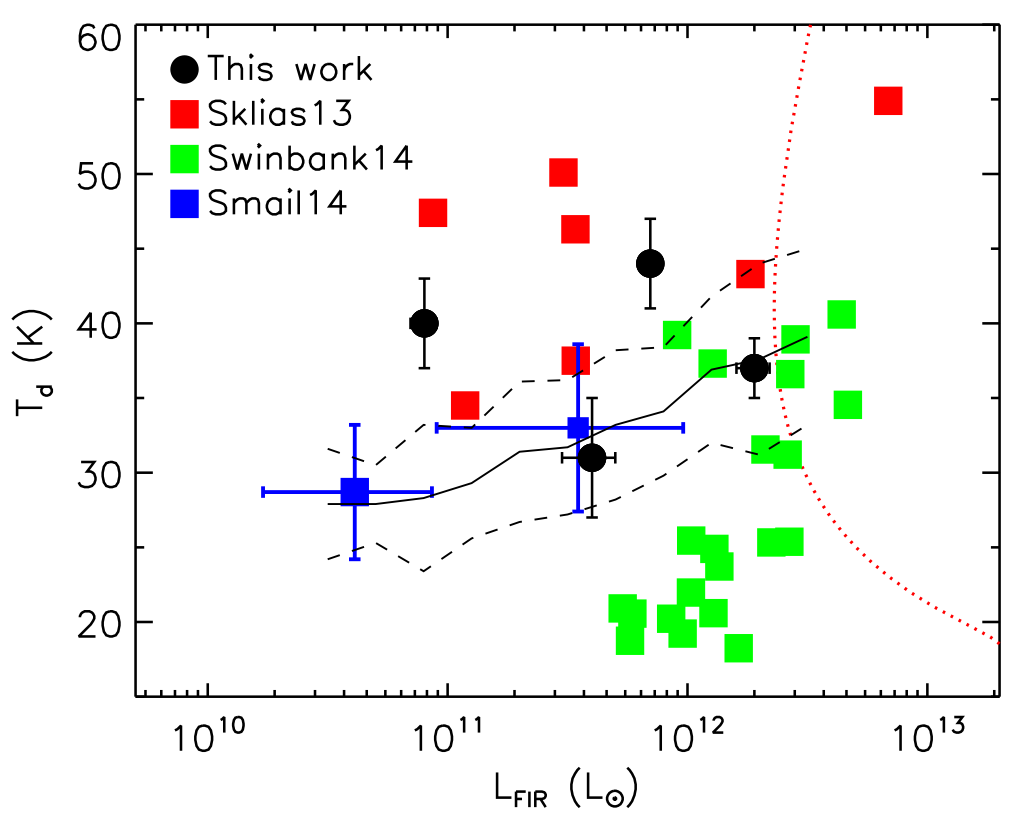}
\end{center}
\caption{Dust temperature versus far-IR luminosity for several samples of galaxies.  The solid line shows the trend found by \protect\cite{symeonidis} using {\it Herschel} for $z\sim$0--1.5 galaxies, with the dashed lines showing the dispersion of the sample.  The green squares are the LESS SMGs followed up by \protect\cite{swinbank} with ALMA and {\it Herschel}, with $z\sim$1-6.  The blue squares are the results of stacking on narrow-band [O{\sc ii}] emitters (left) and MIPS+radio sources not detected in SPIRE/SCUBA-2 (right) for a $z=1.6$ cluster (Smail et al. in press).  The red squares are a sample of lensed SMGs discoverd with {\it Herschel} \protect\citep{sklias} with $z\sim$1.5-3.  The black circles are the four $z\sim2.9$ group galaxies that compose the submm arc of MS\,0451.6$-$0305.  Both \protect\cite{swinbank} and \protect\cite{symeonidis} found that high-$z$ galaxies are on average cooler than the $z=0$ relation, while \protect\cite{sklias} and our results report warmer than average results for high-$z$ galaxies.  The dotted red line represents the SPIRE 250\,$\mu$m detection limit as a function of dust temperature for $z=2.9$ galaxies, illustrating the usefulness of gravitational lensing, to push to fainter objects, when studying high-$z$ SMGs.}
\label{TvsL}
\end{figure}

The SED fits within our model allow us to investigate the physical conditions of each component of the submm arc.  Fig.~\ref{TvsL} plots $T_\mathrm{d}$ verus $L_{\mathrm{IR}}$ for the four $z\sim2.9$ galaxies constrained by our model with trends and data found by \cite{symeonidis}, \cite{swinbank}, \cite{sklias}, and Smail et al. (in press).  As described in \cite{symeonidis}, studying the relation between these two quantities gives insight into the nature of star-formation within galaxies: a flat relation with $T_{\mathrm{d}}=\mathrm{constant}$ implies that star formation regions become more extended when increasing far-IR luminosity, while something close to the Stefan-Boltzmann law, $L_{\mathrm{IR}}\propto T^4$, would imply constant star formation region size (for optically thick star-forming clouds).  \cite{symeonidis} used {\it Herschel} SPIRE and PACS to probe this relation and found the trend plotted as a solid black line in Fig.~\ref{TvsL}, with dashed lines showing the dispersion.  When comparing low and high redshift galaxies, they found that the later were up to 10\,K cooler than their low redshift counterparts, suggesting evolution with redshift towards more extended star-forming regions in the early universe.  \cite{swinbank} found a similar trend with high redshift galaxies being on average 2--3\,K colder than low redshift galaxies.  Smail et al. (in press) found that stacking on narrow-band [O{\sc ii}] emitters and MIPS+radio sources within a $z=1.6$ cluster (intrinsically faint sources) found no evidence of evolution, although their direct detections with SPIRE and SCUBA-2 (thus intrinsically luminous sources) were also found to be cooler in temperature.  A recent study by \cite{sklias}, used gravitational lensing to examine intrinsically fainter galaxies at high redshifts.  Although limited by small number statistics, they found the opposite trend for high redshift galaxies.  When adding the four $z\sim2.9$ galaxies constrained by our model, our results appear to support those found by \cite{sklias}.  This suggests that selection effects and/or biases are present here.  

As has been pointed out before \citep[e.g.][]{chapman2005,chapin2011} selection effects can be extremely important when studying the correlation between $T_\mathrm{d}$ and $L_{\mathrm{IR}}$.  The \cite{swinbank} sample of SMGs were selected at 870\,$\mu$m and thus may be biased towards lower dust temperatures, and those of \cite{sklias} were formally selected at 160\,$\mu$m, and thus could be biased towards warmer dust temperatures.  It should be noted that the submm arc in MS\,0451.6$-$0305 was first discovered at 850\,$\mu$m \citep{first} and therefore unlikely to be biased towards the warmer dust temperatures that we find.

In addition to the selection biases inherent in focussing on a single distinctive object, there are also a number of systematic uncertainties that could be present in our modelling approach.  Most importantly, we have fixed the amplification factors for the galaxy images.  Any errors in amplification can affect our results in several ways.  For example, since the contributions to the submm arc from Galaxies 7 and 8 (See Fig.~\ref{fig:components}) are mostly point-like, then any uncertainty in amplification predominantly affects their measured far-IR luminosities and SFRs.  This is especially true for Galaxy 7, because two of its images lie very close to the critical line, and thus its amplification is highly sensitive to any offset between optical and submm components of the galaxy.  
The uncertainty in relative amplification between galaxy images likely affects which galaxies are preferred by the data.  For example, the images of Galaxies 5 and 6 are spatially very close, thus the different relative amplifications between their respective images probably contributes to Galaxy 6 being preferred by the model fits.

It is also possible that the simple SED model we have adopted may not accurately approximate the true SEDs of the galaxies in the lensed system.  The dust emissivity, $\beta$, is known to be partially degenerate with dust temperature and we have fixed it to a nominal value of 1.5, thus the uncertainties reported for dust temperatures are likely too small.
Furthermore, although this newer {\it HST} data is both deeper and at a longer wavelength, it is possible that we are missing fainter group members, as was the case in previous studies \citep{Borys2004,Berciano2010}.
If any of the galaxies are not at $z\sim2.9$, their reported far-IR luminosities and thus SFRs will be affected, since the distances to the galaxies are used in these calculations.  This is especially true for Galaxy 8, as we have no constraints on its actual redshift and our analysis has assumed it to be part of the $z\sim2.9$ group.

Despite these reservations, the model we have adopted appears to provide a reasonably good fit to the data across a wide range of wavelengths.  Higher resolution submm data would be needed to further investigate the nature of the $z\sim2.9$ galaxy group.

\section{Conclusions}
\label{conclusions}

With our new modelling approach, we have overcome the confused nature of this complex system by fully exploiting the differential amplification across the galaxy group and the multiple imaging caused by the strong gravitational lensing.  This has allowed us to tackle the challenge of disentangling and fitting SEDs to multiple components of the submm arc.  We have shown that the submm arc is predominantly generated by four of the seven galaxies that probably comprise a group at a redshift of $z\sim2.9$, with star-formation likely triggered by the galaxies undergoing a merger. It thus appears that no hidden region of dust-enshrouded star formation \cite[as postulated by][]{Berciano2010} is required to explain the morphology of the submm arc.  This method also demonstrates the power of a broad multi-wavelength approach to fully understanding the nature of the submm arc: {\it HST} imaging gives us the priors on galaxy positions, as well as providing the constraints for the lensing model; {\it Herschel\/} samples the peak of the far-IR SED, as well as providing the high-resolution far-IR imaging at 160\,$\mu$m; and SCUBA-2 850\,$\mu$m data samples the long wavelength portion of the FIR SED at a resolution that closely matches that of the 160\,$\mu$m imaging.

This is a very unique system that gives us a glimpse into the formation of structure and stars in the early Universe, and no other submm lens discovered to date can match the number of separate galaxies lensed from the same redshift.  Spectroscopy and high-resolution follow-up with new interferometer observatories will be the key to confirming and unravelling the nature of this high-$z$ merging galaxy group.

\section*{Acknowledgements}

This research has been supported by the Natural Sciences and Engineering Research Council of Canada.
IRS acknowledges support from STFC (ST/I001573/1), a Leverhulme Fellowship, the ERC Advanced Investigator program DUSTYGAL and a Royal Society/Wolfson merit award.
The James Clerk Maxwell Telescope is operated by the Joint Astronomy Centre on behalf of the Science and Technology Facilities Council of the United Kingdom, the National Research Council of Canada, and (until 31 March 2013) the Netherlands Organisation for Scientific Research. Additional funds for the construction of SCUBA-2 were provided by the Canada Foundation for Innovation.
SPIRE has been developed by a consortium of institutes led by Cardiff University (UK) and including Univ. Lethbridge (Canada); NAOC (China); CEA, LAM (France); IFSI, Univ. Padua (Italy); IAC (Spain); Stockholm Observatory (Sweden); Imperial College London, RAL, UCL-MSSL, UKATC, Univ. Sussex (UK); and Caltech, JPL, NHSC, Univ. Colorado (USA). This development has been supported by national funding agencies: CSA (Canada); NAOC (China); CEA, CNES, CNRS (France); ASI (Italy); MCINN (Spain); SNSB (Sweden); STFC, UKSA (UK); and NASA (USA).
PACS has been developed by a consortium of institutes led by MPE (Germany) and including UVIE (Austria); KU Leuven, CSL, IMEC (Belgium); CEA, LAM (France); MPIA (Germany); INAF-IFSI/OAA/OAP/OAT, LENS, SISSA (Italy); IAC (Spain). This development has been supported by the funding agencies BMVIT (Austria), ESA-PRODEX (Belgium), CEA/CNES (France), DLR (Germany), ASI/INAF (Italy), and CICYT/MCYT (Spain).
This research has made use of data from the HerMES project (\url{http://hermes.sussex.ac.uk/}), a Herschel Key Programme utilising Guaranteed Time from the SPIRE instrument team, ESAC scientists and a mission scientist. 
The HerMES data were accessed through the HeDaM database (\url{http://hedam.oamp.fr}) operated by CeSAM and hosted by the Laboratoire d'Astrophysique de Marseille.
This research used the facilities of the Canadian Astronomy Data Centre operated by the National Research Council of Canada with the support of the Canadian Space Agency. 
Based on observations made with the NASA/ESA Hubble Space Telescope, obtained from the data archive at the Space Telescope Institute. STScI is operated by the association of Universities for Research in Astronomy, Inc. under the NASA contract NAS 5-26555.
This work is based [in part] on observations made with the Spitzer Space Telescope, which is operated by the Jet Propulsion Laboratory, California Institute of Technology under a contract with NASA. Support for this work was provided by NASA.
The Dark Cosmology Centre is funded by the Danish National Research Council.

\bibliography{references}

\begin{table*}
\begin{center}
\caption{List of images for the eight high-$z$ galaxies, as well as one low redshift interloper at $z=0.157$.  The galaxy IDs denote each galaxy, as shown in Fig.~\ref{fig:comparisons}, and the letters indicate the multiple images of each galaxy (with $a$ being the most Northern images in each case, and $c$ being the most southern images).  The position of image 4.b, as inferred from the lensing model, is obscured by foreground cluster galaxies.  The amplification factors are derived from the {\sc Lenstool} modelling in \S~\ref{lenstoolmodel}.  The redshift of Galaxy 8 is unknown, but has similar colours to the other high-redshift multiply imaged galaxies, a disturbed morphology, and was found to be important for reproducing the SW extension of the submm arc, thus we assume a nominal redshift of 2.9. The superscript letters on the redshifts denote the method by which they were derived: {\it a} for redshifts derived from the lensing model, {\it b} for a spectroscopic redshift, and {\it c} for a nominally chosen value.  The reported magnitudes are AB magnitudes.}
\label{images}
\end{center}
\centering
\begin{tabular}{cccccccc} \hline
Galaxy ID  & R.A. & Dec. & F160W & F110W & Amplification & Redshift & Notes \\ 
& J2000 & J2000 & & & & & \\\hline
1.$a$ & 04:54:13.42 & $-$3:00:43.0 & $21.94\pm0.01$ & $23.26\pm0.01$ & $3.80\pm0.06$ & $3.11\pm0.03^a$ & B\citep{Takata2003}\\
1.$b$ & 04:54:12.65 & $-$3:01:16.5 & $20.91\pm0.01$ & $22.27\pm0.01$ & $20\pm1\phantom{0}$ & & C\citep{Takata2003}\\
1.$c$ & 04:54:12.17 & $-$3:01:21.4 & $21.86\pm0.01$ & $23.18\pm0.01$ & $7.3\pm0.1$ & & D\citep{Takata2003}\\
2.$a$ & 04:54:13.15 & $-$3:00:38.4 & $24.15\pm0.03$ & $24.74\pm0.05$ & $2.86\pm0.04$ & $2.91\pm0.04^a$ &\\
2.$b$ & 04:54:12.58 & $-$3:01:11.9 & $23.62\pm0.03$ & $24.25\pm0.05$ & $8.1\pm0.4$ & & \\
2.$c$ & 04:54:11.79 & $-$3:01:20.2 & $22.88\pm0.02$ & $23.85\pm0.04$ & $6.1\pm0.1$ & & \\
3.$a$ & 04:54:13.04 & $-$3:00:39.2 & $24.98\pm0.04$ & $26.28\pm0.07$ & $3.19\pm0.05$ & $2.94\pm0.04^a$ &\\
3.$b$ & 04:54:12.68 & $-$3:01:09.1 & $23.27\pm0.02$ & $24.09\pm0.05$ & $2.98\pm0.05$ &  &\\
3.$c$ & 04:54:11.46 & $-$3:01:21.7 & $24.27\pm0.04$ & $25.49\pm0.06$ & $4.31\pm0.08$ &  &\\
4.$a$ & 04:54:12.82 & $-$3:00:39.3 & $24.82\pm0.05$ & $26.39\pm0.08$ & $3.57\pm0.06$ & $2.94\pm0.04^a$ &\\
4.$b$ & {\it 04:54:12.53} & {\it $-$3:01:04.5} & $26.64\pm0.07$ & $27.50\pm0.09$ & $6.2\pm0.2$ & & Lensing model position\\
4.$c$ & 04:54:11.03 & $-$3:01:22.4 & $24.70\pm0.05$ & $25.90\pm0.07$ & $3.36\pm0.06$ & & \\
5.$a$ & 04:54:12.81 & $-$3:00:44.4 & $21.73\pm0.01$ & $23.51\pm0.01$ & $5.3\pm0.1$ & $2.89\pm0.03^a$ & ERO-B\citep{Borys2004}\\
5.$b$ & 04:54:12.69 & $-$3:01:01.5 & $21.81\pm0.01$ & $23.47\pm0.01$ & $6.4\pm0.1$ &  & ERO-B\citep{Borys2004}\\
5.$c$ & 04:54:10.93 & $-$3:01:24.6 & $21.97\pm0.01$ & $23.78\pm0.02$ & $2.89\pm0.04$ &  & ERO-B\citep{Borys2004}\\
6.$a$ & 04:54:12.81 & $-$3:00:47.5 & $22.62\pm0.02$ & $24.55\pm0.06$ & $8.2\pm0.2$ & $2.86\pm0.03^a$ & ERO-C\citep{Borys2004}\\
6.$b$ & 04:54:12.72 & $-$3:00:59.6 & $24.41\pm0.04$ & $26.60\pm0.15$ & $4.98\pm0.08$ &  & ERO-C\citep{Borys2004}\\
6.$c$ & 04:54:10.88 & $-$3:01:25.8 & $22.85\pm0.02$ & $24.67\pm0.09$ & $2.76\pm0.04$ &  & ERO-C\citep{Borys2004}\\
7.$a$ & 04:54:12.95 & $-$3:00:54.8 & $21.80\pm0.01$ & $22.26\pm0.01$ & $33\pm2\phantom{0}$ & $2.911\pm0.003^b$ & LBG\citep{Borys2004}\\
7.$b$ & 04:54:12.93 & $-$3:00:57.5 & $22.29\pm0.01$ & $22.76\pm0.01$ & $45\pm3\phantom{0}$ &  & LBG\citep{Borys2004}\\
7.$c$ & 04:54:11.11 & $-$3:01:26.6 & $23.66\pm0.02$ & $24.23\pm0.03$ & $2.87\pm0.04$ &  & LBG\citep{Borys2004}\\
8 & 04:54:10.55 & $-$3:01:27.3 & $22.77\pm0.02$ & $23.50\pm0.03$ & $1.73\pm0.04$ & $2.9^c$ & Singly imaged\\
9 & 04:54:12.85 & $-$3:01:09.1 & $18.91\pm0.01$ & $19.19\pm0.02$ & -- & $0.15719^b$ & foreground galaxy\\ \hline
\end{tabular}
\end{table*}

\begin{table*}
\centering
\begin{center}
\caption{Lensing-amplification-corrected results from the model.  The total $L_{\mathrm{FIR}}$ for the $z\sim2.9$ galaxy group is $(3.1\pm0.3) \times 10^{12}\mathrm{L}_\odot$, which gives a SFR of $(450\pm50)$ M$_\odot$yr$^{-1}$.  The 95$^\mathrm{th}$ percentile upper limits are given for galaxies not found to be contributing to the submm arc.  Note that Galaxy 9 is a foreground galaxy at $z=0.157$ and is therefore not lensed.}
\label{results}
\end{center}
\begin{tabular}{cccccccccc} \hline
Gal & $T_{\mathrm{d}}$ & $L_{\mathrm{FIR}}$  & SFR & $S_{160}$ & $S_{250}$ & $S_{350}$ & $S_{500}$ & $S_{450}$ & $S_{850}$ \\ 
ID & ($\mathrm{K}$) & ($\mathrm{L}_\odot$) &  ($\mathrm{M}_\odot \mathrm{yr}^{-1}$)  &(mJy)&(mJy)&(mJy)&(mJy)&(mJy)&(mJy)\\ \hline
1 & $-$ & $<8.2 \times 10^{9}$ & $<1.3$ & $-$& $-$& $-$& $-$& $-$& $-$\\
2 & $44\pm3$ & $(6.7\pm0.6) \times 10^{11}$ & $99\pm9$ & $0.94\pm0.10$ & $1.9\pm0.2$ & $1.9\pm0.2$ & $1.2\pm0.2$ & $1.4\pm0.2$ & $0.31\pm0.06$\\
3 & $-$ & $<1.5 \times 10^{11}$ & $<23$  & $-$& $-$& $-$& $-$& $-$& $-$\\
4 & $-$ & $<3.3 \times 10^{11}$ & $<50$  & $-$& $-$& $-$& $-$& $-$& $-$\\
5 & $-$ & $<2.0 \times 10^{11}$ & $<35$  & $-$& $-$& $-$& $-$& $-$& $-$\\
6 & $31\pm4$ & $(3.6\pm0.9) \times 10^{11}$ & $53\pm14$ & $0.31\pm0.12$ & $0.7\pm0.3$ & $1.3\pm0.4$ & $1.5\pm0.4$ & $1.5\pm0.4$ & $0.7\pm0.2$\\
7 & $40\pm3$ & $(7.5\pm1.0) \times 10^{10}$ & $11\pm2$ & $0.10\pm0.02$ & $0.22\pm0.04$ & $0.23\pm0.06$ & $0.16\pm0.05$ & $0.18\pm0.04$ & $0.04\pm0.02$\\ 
8 & $37\pm2$ & $(1.9\pm0.3) \times 10^{12}$ & $290\pm40$ & $2.5\pm0.4$ & $5.6\pm0.8$ & $6.9\pm1.0$ & $5.1\pm0.9$ & $5.8\pm1.0$ & $1.5\pm0.3$\\ 
\noalign{\smallskip}
9 & \it{$17\pm9$} & $(1.7\pm0.5) \times 10^{9}$ & $0.25\pm0.07$ & $3.6\pm1.0$ & $2.5\pm0.8$ & $1.3\pm0.5$ & $0.5\pm0.2$ & $0.6\pm0.2$ & $0.08\pm0.04$\\ 
\hline
\end{tabular}
\end{table*}

\end{document}